\documentclass{article}
\usepackage{spconf,amsmath,graphicx,amsfonts,xcolor,url,booktabs,multirow,subcaption,comment,siunitx}
\newcommand{\spcell}[2][c]{%
  \begin{tabular}[#1]{@{}c@{}}#2\end{tabular}}
  
\title{A Memory Augmented Architecture for Continuous\\Speaker Identification in Meetings}

\name{Nikolaos Flemotomos$^{1}$\sthanks{Nikolaos Flemotomos performed this work
	while at Microsoft.}, Dimitrios Dimitriadis$^{2}$}
\address{$^{1}$ Signal Analysis and Interpretation Lab, University of Southern California, Los Angeles, CA, USA\\
$^{2}$ Speech and Dialog Research Group, Microsoft, Redmond, WA, USA}

\begin{document}
\ninept
\maketitle
\begin{abstract}
We introduce and analyze a novel approach to the problem of speaker identification in multi-party recorded meetings. Given a speech segment and a set of available candidate profiles, we propose a novel data-driven way to model the distance relations between them, aiming at  identifying the speaker label corresponding to that segment. To achieve this we employ a recurrent, memory-based architecture, since this class of neural networks has been shown to yield advanced performance in problems requiring relational reasoning. The proposed encoding of distance relations is shown to outperform  traditional distance metrics, such as the cosine distance. Additional improvements are reported when the temporal continuity of the audio signals and the speaker changes is modeled in. In this paper, we have evaluated our method in two different tasks, i.e. scripted and real-world business meeting scenarios, where we report a relative reduction in speaker error rate of $39.28\%$ and $51.84\%$, respectively, compared to the baseline.

\end{abstract}


\section{Introduction}
\label{sec:intro}
Speaker identification is the task of determining the identity of the person uttering a particular phrase, assuming a finite set of pre-enrolled speakers is given \cite{hansen2015speaker}. 
Applying a continuous automatic speaker identification system on 
recorded meetings with multiple participants can significantly affect the performance of several subtasks of the meeting analytics suite. For instance, correctly identifying the active speaker is an essential component for rich meeting transcriptions (Speaker-Attributed Automatic Speech Recognition - SA-ASR)~\cite{fiscus2007rich, yoshioka2019meeting}, speaker tracking~\cite{bonastre2000speaker}, action item generation~\cite{mcgregor2017more}, or speaker adaptation for more reliable ASR outputs~\cite{mimura2011fast}. The main difference of the investigated task with speaker diarization is the use of speaker profiles, since the meeting participants are known in advance~\cite{biagetti2016robust}, i.e. the number of speakers and their acoustic identities are provided.

\begin{figure}[ht]
  \centering
  \includegraphics[width=.9\columnwidth]{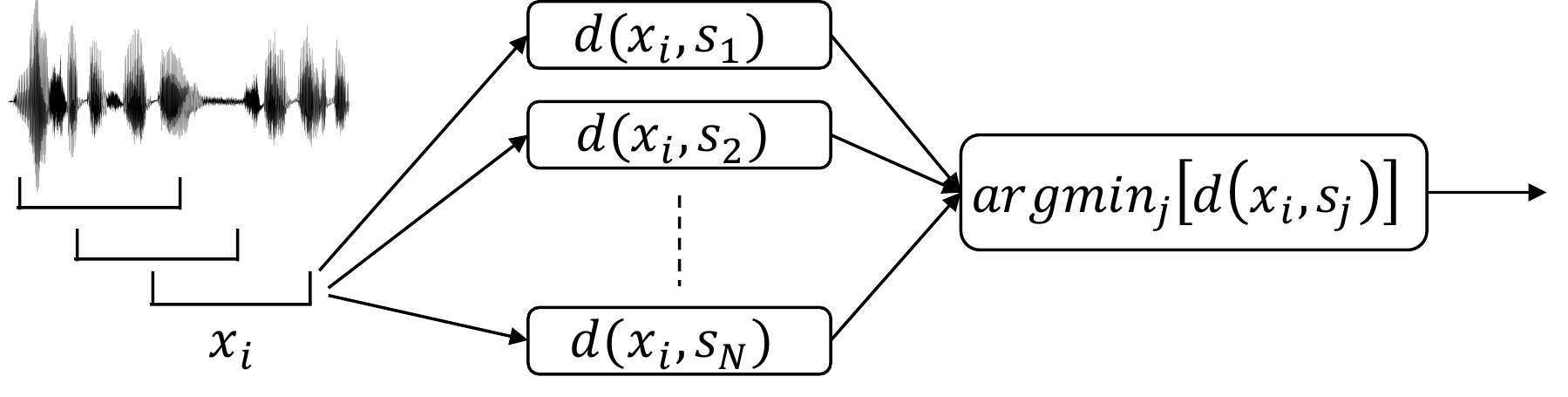}
  \caption{State-of-the-art continuous speaker identification system: The speech signal is segmented uniformly and each segment $x_i$ is compared against all the available speaker profiles $\{s_j\}_{j=1}^N$ according to a metric $d(\cdot,\cdot)$. The speaker label assigned to each $x_i$ minimizes this metric.}
  \label{fig:baseline_nthfar}
\end{figure}

For speaker attribution tasks, an enrollment phase is required. During that phase, sample audio from the participants is collected and the target speaker profiles (or identities) are produced. Continuous speaker identification can be thought of as a two-step problem, since it comprises a  segmentation and a classification phase. Initially, the audio signal is segmented either uniformly~\cite{zajic2016investigation} or based on some estimated speaker change points~\cite{yoshioka2019meeting}. These segments are now assumed speaker-homogeneous\footnote{A single speaker is present.}. Speech embeddings of each segment are extracted and then, compared against all the available speaker profiles. By minimizing a particular distance metric (as described below), the most suitable speaker label is assigned to the segment~\cite{yoshioka2019meeting}.
Initially, the extracted speaker-specific features/embeddings were based on the i-vectors~\cite{dehak2010front}, but lately bottleneck representations from deep learning architectures such as x-vectors~\cite{snyder2018x} are used. The final decision relies either on the cosine~\cite{dehak2010front,yoshioka2019meeting} or the PLDA \cite{garcia2011analysis} distance. The overall process is illustrated in \mbox{Fig. \ref{fig:baseline_nthfar}}.

This approach poses some potential problems. First, uniform segmentation introduces a trade-off decision concerning the segment length: segments need to be sufficiently short to safely assume that they do not contain multiple speakers but at the same time it is necessary to capture enough information to extract a meaningful speaker representation. 
Further, the speaker embeddings are usually extracted from a network trained to distinguish speakers amongst thousands of candidates~\cite{snyder2018x}. However, since a small number of participants is involved in an interactive meeting setup, a different level of granularity in the speaker space is required. 
Also, the distance metric used is often heuristic or depends on certain assumptions which do not necessarily hold, e.g., assuming Gaussianity in the case of  the PLDA distance \cite{garcia2011analysis}. 
Finally, the audio chunks are treated independently and any temporal information about the past and future is simply ignored.

In this work, a data-driven, memory-based approach is proposed that can address some of the aforementioned challenges. Data-driven techniques
perform remarkably well on a wide variety of tasks~\cite{lecun2015deep}; traditional architectures, though, may fail when the problem involves relational information between observations~\cite{santoro2018relational}. Speaker identification can be seen as a member of this class of tasks, since the final decision depends on the distance relations between speech segments and speaker profiles. Herein, the Memory-Augmented Neural Networks (MANNs)~\cite{sukhbaatar2015end,graves2014neural}
are proposed bridging the performance gap.
Based on the success of MANNs on several problems requiring relational reasoning
and specifically using the  Relational Memory Core (RMC) \cite{santoro2018relational}, we build a memory-based network for the task of continuous speaker identification in meeting scenarios. While compared to the baseline approach, we show consistent improvements in performance.  

\section{Method}
\label{sec:method}
\subsection{MANNs and RMC}
\begin{figure}[ht]
  \centering
  \includegraphics[width=.9\columnwidth]{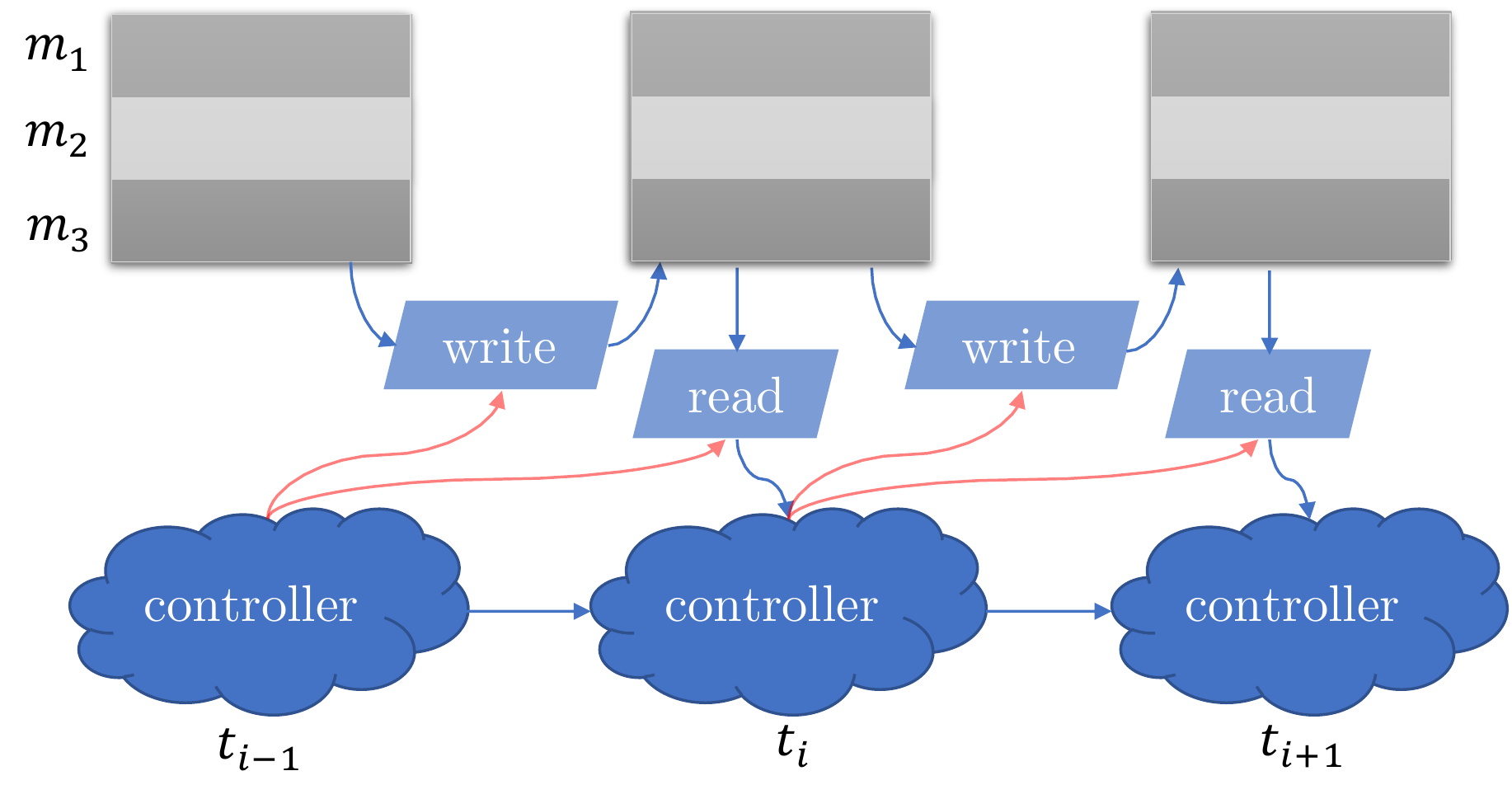}
  \caption{Graphical illustration of how a MANN operates through time. The memory matrix $\mathbf{M}$ here consists of 3 memory slots $\{m_i\}_{i=1}^3$.}
  \label{fig:mann}
\end{figure}

The main concept of MANNs is augmenting a recurrent neural network with a memory matrix $\mathbf{M}\in\mathbb{R}^{Q\times P}$ consisting of $Q$ $P$-dimensional memory slots. The main architecture, called the controller, decides how to update the memory through attention mechanisms using read and write heads, as shown in Fig. \ref{fig:mann}. 
The entire system is differentiable, meaning that it can learn a task-specific organization of the memory in a supervised manner through gradient descent \cite{graves2016hybrid}.

Several implementations of this class of networks have been proposed in the literature (e.g. \cite{graves2014neural,pham2018relational}). For our task, the framework introduced in \cite{santoro2018relational} is used, where the controller of the network is embedded into a Long Short-Term Memory (LSTM) cell and is called Relational Memory Core (RMC). RMC controls the memory updates through a self-attention mechanism \cite{vaswani2017attention} in a way that the memory matrix dimensions remain constant. Each time a new observation is presented, self-attention allows each memory to attend to that observation, as well as to all the other memories, before been updated, so that cross-memory relations are encoded.   

\subsection{RMC-based architecture for speaker identification}
\label{subsec:rmc_id}
\begin{figure}[ht]
  \centering
  \includegraphics[width=.92\columnwidth]{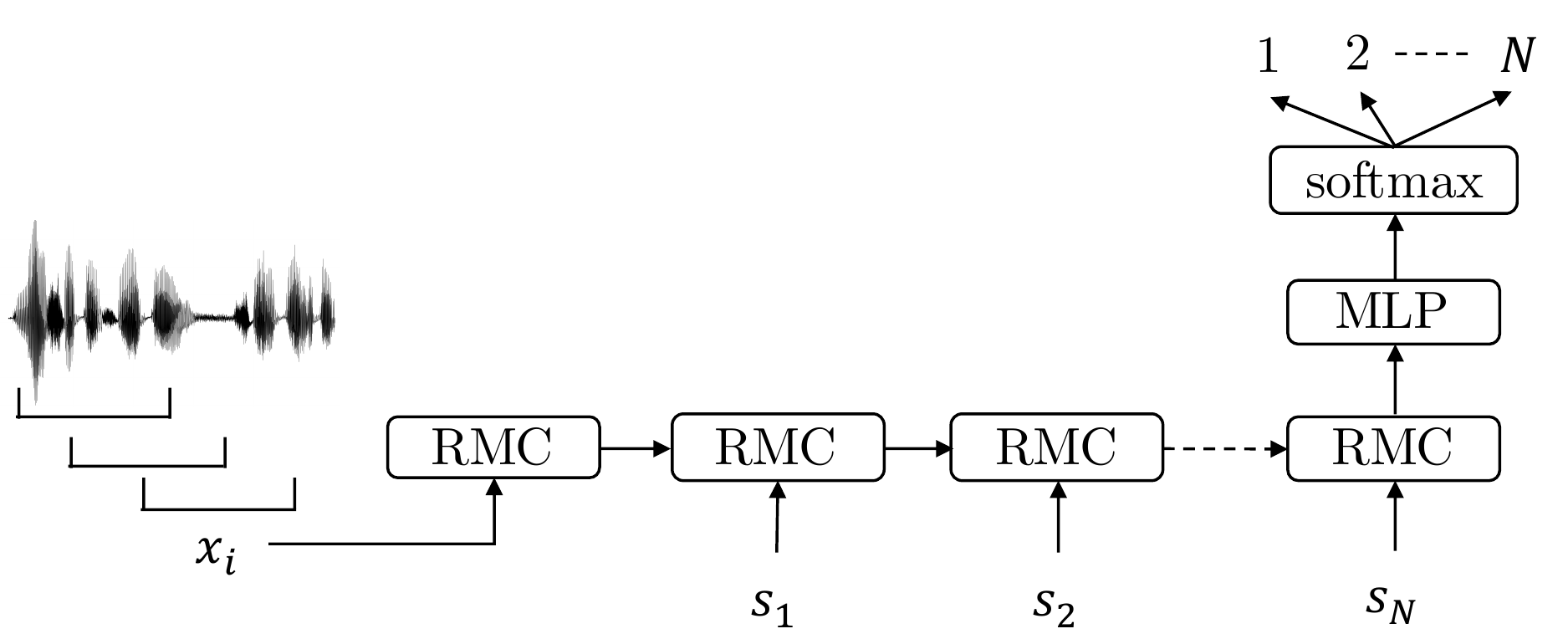}
  \caption{Unrolled recurrent network for continuous speaker identification. $x_i$ represents a speech segment and $\{s_j\}_{j=1}^N$ are the profiles of the $N$ speakers appearing in the recording.}
  \label{fig:proposed_nthfar}
\end{figure}

RMC-based networks have shown good performance on several problems \cite{santoro2018relational}, including the ``$n^\text{th}$ farthest task'', where the goal is to find the $n^\text{th}$ farthest vector from the $m^\text{th}$ element in a given sequence of vectors. Assume we are given a particular audio segment of a meeting $x_i$ and a set of profiles $\{s_j\}_{j=1}^N$ corresponding to the $N$ participants in the meeting. Under the $n^\text{th}$ farthest task notation, we construct the sequence $S_i = \{x_i,s_1,s_2,\cdots,s_N\}$ and we view speaker identification as the problem of finding the closest element to $x_i$ in the sequence $S_i$. In more detail, a  sequence of vectors $S_i$ for each audio segment $x_i$ of the input audio signal is passed through an RMC-based recurrent layer, as in Fig. \ref{fig:proposed_nthfar}. The output of this layer goes through a fully-connected Multilayer Perceptron (MLP) with a softmax inference layer returning the label $l_i\in\{1,2,\cdots,N\}$; that is the one maximizing the probability $\mathbb{P}\left[l_i=j|x_i,\{s_j\}_{j=1}^N\right]$. Intuitively, the RMC projects each element of the input sequence onto the ``memory space'' and the network learns some local data-driven distance metric, sorts the resulting distances, and finds the profile that yields the minimum distance.

Note that $N$ is a prefixed maximum number of speakers within a meeting that the network can handle. Given a sequence with \mbox{$\tilde{N}<N$} 
profiles,  the remaining $N-\tilde{N}$ outputs of the softmax layer are expected to be close to zero. To that end,  the network is trained with variable length sequences, providing training examples with all the expected numbers of participants in a meeting. 

As shown in Section \ref{sec:res}, the proposed architecture needs a large number of training sequences, containing many speaker profiles. This may cause overfitting when the in-domain meeting data are limited.
Instead, an out-of-domain dataset can be used to construct speaker profiles and generate sequences of speech segments and random profiles\footnote{The only constraint being that the ground truth profile corresponding to a segment should be included in the constructed sequence.}. 
In case we have multiple training sequences with the same speaker profiles (e.g. we use long real-world meeting data for training), we need to be cautious so that the network learns distance relations and not specific positions of the profiles within the sequences. For that reason, the profiles $\{s_j\}_{j=1}^N$ of each sequence are randomly permuted during training.

\subsection{Incorporating temporal information}
As mentioned in the Introduction, one of the challenges of the traditional approaches is the lack of temporal information. However, adding temporal context in our proposed approach is quite straightforward, by constructing the input sequences while including the information from neighboring segments. For example, to identify the speaker profile that corresponds to $x_i$ with a temporal context of $1$, we would use the sequence $\{x_{i-1}, x_i, x_{i+1}, s_1, s_2, \cdots, s_N\}$, where $\{s_j\}_{j=1}^N$ are the candidate speaker profiles.

Temporal continuity is a well-known issue at the decision level as well since it is highly improbable that isolated short segments correspond to some speaker in  the middle of an utterance assigned to another speaker. Within the diarization community, this is often addressed through Viterbi or Variational Bayes resegmentation \cite{kenny2010diarization, sell2015diarization}. For this work we apply a simple smoothing of the trajectory of the predicted speaker labels through median filtering. We note that we could similarly introduce a Hidden Markov Model (HMM) for trajectory smoothing.

\section{Datasets}
\label{sec:data}
The publicly available AMI corpus \cite{carletta2005ami} consists of meeting data, either occurring naturally or following a scripted scenario. For our experiments we use the scripted meetings with $4$ speakers each, with both close-talk and far-field audio available, giving us $31$ scenarios. Each scenario consists of $4$ meetings happening throughout a day: the first set of meetings is used for the speaker profile estimation (totally \SI{8.0}{h} ignoring silence), the second and third for training (\SI{35.5}{h}), and the fourth for evaluation (\SI{17.1}{h}). We refer to this evaluation set as the \texttt{seen} one, since the speakers are seen during training. An additional evaluation set with \texttt{unseen} speakers is used, consisting of $6$ meetings (\SI{4.1}{h}) -- while $6$ more meetings (\SI{3.8}{h}) with the same speakers are used for their profile extraction. To resemble real-world conditions, far-field audio is used for training and evaluation, while the profile estimation is based on the close-talk microphones, as would normally be done during the enrollment phase. 

To introduce more speaker variability during training, we additionally use data from VoxCeleb 1 and 2 \cite{nagrani2017voxceleb, chung2018voxceleb2}.
We have kept all the speakers with more than $6$ utterances each, resulting to a subset containing  $6$,$490$ speakers. For each speaker, $3$ utterances are randomly selected for profile estimation (totalling \SI{383.3}{h}) and the rest (\SI{2184.9}{h}) is used for training.  

The method is also evaluated on $9$ real meetings recorded within Microsoft with a circular microphone array \cite{yoshioka2019advances}. The seven channels of the array are combined through a differential beamformer, processing the beam output with the highest energy. The number of speakers in those meetings (total length of \SI{4.6}{h}) ranges from $4$ to $15$ and all the speakers had been already enrolled while reading short text excerpts with a close-talk microphone.

\section{Experiments and Results}
\label{sec:res}

\subsection{Experimental setup}
We will use the ground truth Voice Activity Detection (VAD) segmentation as provided by the available human-generated transcripts while ignoring the speaker labels. Consecutive speech segments with an in-between silence shorter than \SI{0.5}{sec} are now merged. This segmentation scenario will be noted as the \texttt{oraclevad}, as opposed to the \texttt{oraclespk}, where the initial speaker label info is also included. 
The main difference between the two scenarios is that segments in \texttt{oraclevad} may contain more than one speakers, while each segment in \texttt{oraclespk} contains a single speaker. 
While generating the training and evaluation sequences, a sliding analysis window with length of \SI{1.5}{sec} is used for the x-vector extraction. The window shift is \SI{0.75}{sec} for the AMI and the internal meeting dataset, and \SI{10}{sec} for the VoxCeleb dataset. This process creates $160$K training examples for AMI and $841$K subsegments for VoxCeleb. A \mbox{$512$-dimensional} \mbox{x-vector} is generated per window, using the pretrained VoxCeleb model\footnote{\url{https://kaldi-asr.org/models/m7}} provided by the Kaldi toolkit~\cite{povey2011kaldi}. The x-vectors are decorrelated via an LDA projection (after which we keep $200$ dimensions) and are further mean- and length-normalized~\cite{garcia2011analysis}.
A speaker profile is estimated as the per speaker mean of all the \mbox{x-vectors} estimated on the available speaker-homogeneous segments, in the \texttt{oraclespk} scenario.

The memory matrix has $(N+1)$ $2048$-dimensional memory slots, where $N$ is the maximum number of speakers the network expects to see. The MLP component consists of $4$ fully connected layers of $256$ neurons each. The network is built using TensorFlow~\cite{abadi2016tensorflow} and the Sonnet library\footnote{\url{https://github.com/deepmind/sonnet}}. 

In the case of \texttt{oraclespk} segmentation, the evaluation metric is based on the window-level (subsegment-level) classification accuracy. On the other hand, for the \texttt{oraclevad} segmentation we rely on the Speaker Error Rate (SER) as estimated by the NIST \texttt{md-eval.pl} tool, with a \SI{0.25}{sec}-long collar, while ignoring overlapping speech segments. The system described in \mbox{Fig. \ref{fig:baseline_nthfar}} with cosine distance is used as the baseline.

\subsection{Results on AMI}
\begin{figure}[ht]
  \centering
  \includegraphics[width=.9\columnwidth]{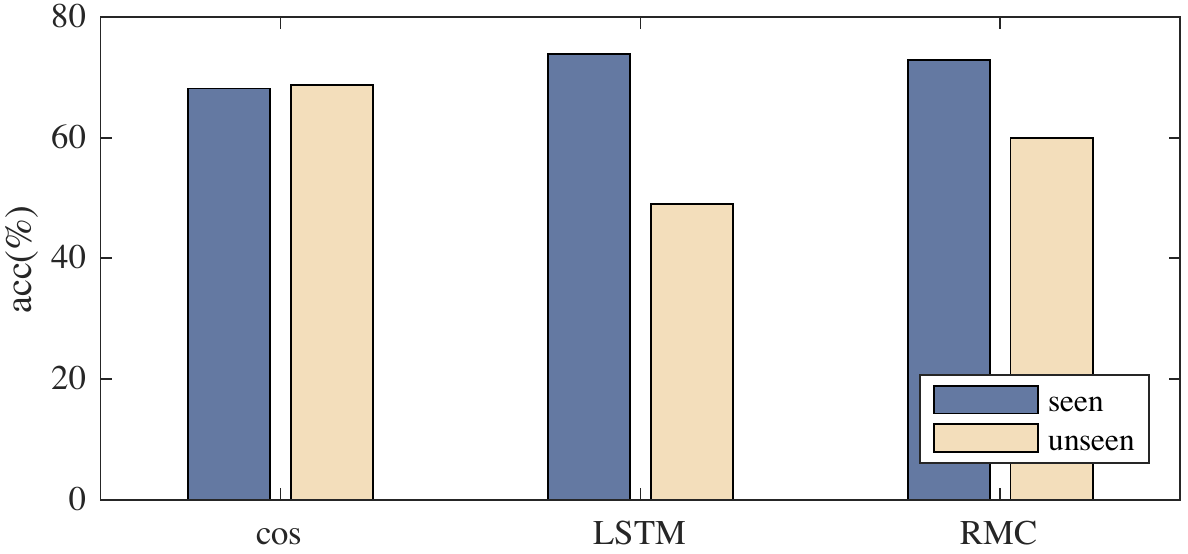}
  \caption{Classification accuracy on AMI for the \texttt{seen} and \texttt{unseen} evaluation sets with \texttt{oraclespk} segmentation. Both the LSTM- and RMC-based networks are trained on AMI.}
  \label{fig:AMI_1st_res}
\end{figure}

First, the system is trained and evaluated on the AMI corpus using the \texttt{oraclespk} segmentation. Apart from the baseline and the RMC-based systems, a third one  replacing the RMC-based layer of \mbox{Fig. \ref{fig:proposed_nthfar}} with an LSTM layer, is also evaluated. The RMC appears capturing the desired distance information better than the LSTM, but both networks are outperformed by the baseline on \texttt{unseen} speakers scenario, as shown in \mbox{Fig. \ref{fig:AMI_1st_res}}. In particular, the performance of the baseline system is practically the same for all testing conditions (i.e., \texttt{seen} vs. \texttt{unseen}). This is expected as the system is based solely on the cosine distances. On the contrary, the behavior of the recurrent networks appears quite different with a big performance gap between the \texttt{seen} and \texttt{unseen} evaluation sets. A possible explanation is that the networks overfit to the training speakers and fail to generalize. Also, the LSTM-based network performs substantially worse in the \texttt{unseen} conditions than the RMC-based. So, henceforth we will only use the latter one.

To avoid the overfitting  during training with the limited number of training speakers in AMI, we decided to use randomly generated training sequences from VoxCeleb, as described in \mbox{Section \ref{subsec:rmc_id}}. 
We note that even when the VoxCeleb training recordings are  distorted by reverberation and noise to simulate the meeting environment, the speaker profiles are always estimated on clean audio. Based on the results  in \mbox{Table \ref{table:AMI_res}}, including  the  VoxCeleb dataset for training, thus incorporating greater variability in the speaker acoustic characteristics,  leads to performance comparable to the baseline system for the \texttt{unseen} evaluation set. Further distortion simulating real-world conditions  boosts significantly the system performance. The addition of temporal context (last row of Table \ref{table:AMI_res}) gives additional substantial improvements.

\begin{table}[ht]
  \centering
	\begin{tabular}{c|c|c}
    	\toprule
        system & training set & acc $(\%)$\\
        \midrule
        cos & -- & $68.68$\\
        \hline
        \multirow{4}{*}{RMC}&AMI & $60.00$ \\
		&VoxCeleb clean & $68.15$ \\
		&VoxCeleb reverb & $70.25$\\
		&VoxCeleb reverb+noise & $71.90$\\
        \hline
        RMC \& context ($\pm 1$) & VoxCeleb reverb+noise & $\mathbf{73.86}$\\
       	\bottomrule
	\end{tabular}    
	\caption{Classification accuracy on AMI for the \texttt{unseen} eval set with \texttt{oraclespk} segmentation, when the system is trained on different training sets, with or without context.}
	\label{table:AMI_res}
\end{table}

Up to this point all sequences for training and evaluating have a fixed number of speakers (equal to~$4$). A more realistic system, used on meetings, should support a variable number of participants. Thus, we now train the system with variable-length sequences, with or without context and in Table~\ref{table:var_len} we report the results when it is still evaluated in AMI ($4$ speakers). As expected from the analysis and results of the previous experiments, adding context leads to consistent performance improvements. In the case when there is not temporal information, an additional observation is that, as the range increases, there is a decreasing trend in the classification accuracy.
However, comparing the columns $1$ and $4$ in Table~\ref{table:var_len}, it seems that adding context not only improves the performance, but also makes the system more robust, when this is trained on larger ranges of sequence lengths.

\begin{table}[ht]
    \centering
        \begin{tabular}{c|c|c|c|c}
        \toprule
        \spcell{training seq\\length} & {$4$ spks} & {$4$-$6$ spks} & {$2$-$9$ spks} & {$4$-$15$ spks}\\
        \midrule
        w/o context & $71.90$ & $71.94$ & $70.84$ & $69.66$\\
        with context & $73.86$ & $73.77$ & $72.67$ & $73.42$ \\
        \bottomrule
        \end{tabular}
    \caption{Classification accuracy $(\%)$ on AMI for the \texttt{unseen} evaluation set with \texttt{oraclespk} segmentation, when the system is trained on VoxCeleb (with added noise and reverbaration), with diffferent ranges of sequence lengths.}
    \label{table:var_len}
\end{table}

\vspace*{-.2cm}
\subsection{Results on internal meetings}
The system trained on VoxCeleb sequences of $4$-$15$ speakers (last column of Table \ref{table:var_len}) is also evaluated on  $9$ internal meetings, where the number of participants varies. Both the \texttt{oraclevad} and the \texttt{oraclespk} segmentation scenarios are investigated. Once again,  the memory-based network yields superior performance compared to the baseline (\mbox{Table \ref{table:MSFT_res}}) especially when temporal information from the neighboring subsegments is added. 

\begin{table}[ht]
	\centering
	\begin{tabular}{c|c|c|c}
	    \toprule
		&{cos} & {RMC} & {RMC \& context}\\
		\midrule
		\spcell{\texttt{oraclevad} -- SER $(\%)$\\\hspace*{.04cm} lower is better} & $20.95$ & $18.56$ & $\mathbf{11.69}$\\
		\hline
		\spcell{\texttt{oraclespk} -- acc $(\%)$\\higher is better} & $70.66$ & $72.51$ & $\mathbf{79.97}$\\
	    \bottomrule
	\end{tabular}
	\caption{System evaluation on the internal Microsoft meetings with different initial segmentation approaches.}
    \label{table:MSFT_res}
\end{table}

The relative performance gain when adding the temporal context to the system is substantially larger, compared to the results on the AMI meetings (Table~\ref{table:var_len}). This behavior can be partly explained by the inherent differences in the acoustic conditions between the two datasets. For instance, about $16\%$ of the speaking time in the \texttt{unseen} AMI evaluation set is overlapping speech, while the corresponding percentage for the internal meetings is about $7\%$. Similar discrepancies were observed with regards to the frequency of speaker change points. As a consequence, it is more probable that neighboring segments of the internal meeting recordings contain information about the same speaker, thus boosting the performance when jointly provided to the network.

\subsection{Smoothing at the decision level}
\begin{figure}[ht]
  \centering
  \begin{subfigure}{.49\linewidth}
  \centering
  \includegraphics[width=\linewidth]{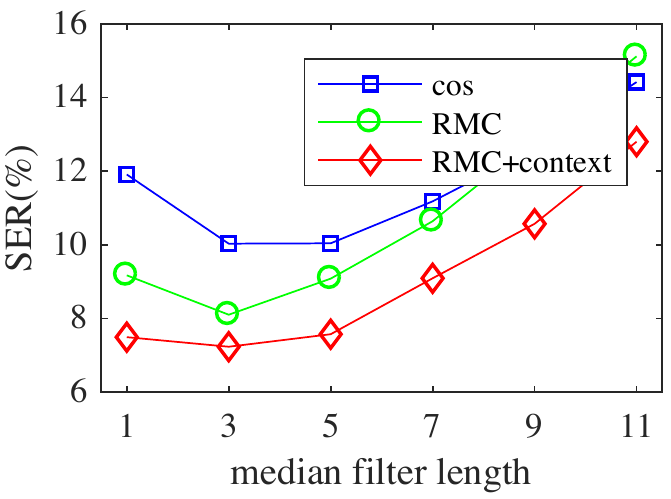}
  \caption{AMI, \texttt{unseen}.}
  \end{subfigure}
  \begin{subfigure}{.49\linewidth}
  \centering
  \includegraphics[width=.97\linewidth]{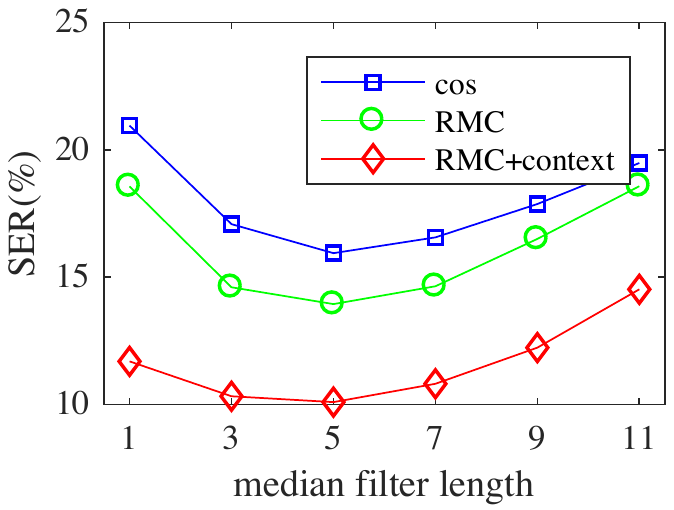}
  \caption{Internal meetings.}
  \end{subfigure}
  \caption{SER as a function of the post-processing median filter length for the two evaluation datasets with \texttt{oraclevad} segmentation. The RMC-based network is trained on sequences of $4$-$15$ speakers.}
  \label{fig:median_filt}
\end{figure}

Finally, we investigate the effect of temporal continuity on the output decisions by introducing median filtering.
It is shown (Fig.~\ref{fig:median_filt}) that a short median filter of a few taps can improve the overall performance for both datasets. Similar patterns are observed for both  the baseline and  the memory-based approach. Even though the results are also improved for the RMC-based network,  trained with temporal context from neighboring segments, the relative improvements are substantially smaller. We have concluded that adding temporal context to the network partially acts like a data-driven smoothing filter. Such a smoothing seems to be more effective than applying a post-processing filter (in our case median filter). This becomes apparent from the fact that the RMC-based network without added context is always outperformed by the network with added temporal information, even if no post-processing trajectory smoothing is applied to the latter (median filter length = 1 tap).

Overall, the minimum SER achieved with the RMC-based architecture including context is $7.23\%$ for the AMI data (median filter length = $3$ taps) and $10.09\%$ for the internal meetings (median filter length = $5$ taps). The cosine-based system without median filtering yields a SER equal to $11.91\%$ and $20.95\%$, respectively, which translates to a relative SER reduction of $39.29\%$ and $51.84\%$.

\vspace*{-.15cm}
\section{Discussion and Future Work} \label{sec:conc}
Herein, we have proposed a system suitable for the task of continuous speaker identification in meeting scenarios. This is based on a recurrent memory network which is capable of modeling the distance relations between observations; that is between speaker profiles and speech segments. 
We evaluated our approach on two corpora featuring conversational business interactions under different conditions. The proposed system yields consistent improvements in performance, when compared to a baseline system depending on the cosine distance metric. We have additionally emphasized the importance of incorporating temporal context both at the feature and the decision level, as well as the beneficial effects of using a training dataset with a large variety of speakers, and with environmental conditions matching the testing conditions, even if artificially. Following our best configuration, we achieved a SER relative reduction of $39.29\%$ for the AMI corpus and $51.84\%$ for the internal Microsoft meetings, when using oracle VAD information. 

A potential extension of the current work will focus on better context modeling -- e.g. incorporating transition probabilities between the various speakers -- and on alternative memory-based architectures, which can generalize to the problem of diarization, capturing the profile information on the fly.


\vfill\pagebreak

\ninept
\bibliographystyle{IEEEbib}
\bibliography{refs}

\end{document}